\documentclass[twoside,twocolumn,english]{revtex4-1}
\usepackage[T1]{fontenc}
\usepackage[latin9]{inputenc}
\usepackage{amsmath}
\usepackage{amssymb}
\usepackage{graphicx}
\usepackage{esint}

\makeatletter

\usepackage{soul}
\usepackage{color}
\usepackage{babel}
\usepackage{calc}

\makeatother

\usepackage{babel}
\begin{document}
\title{Transport and Localization in Quantum Walks on a Random Hierarchy
of Barriers}
\author{Richa Sharma and Stefan Boettcher}
\affiliation{Department of Physics, Emory University, Atlanta, GA 30322; USA}
\begin{abstract}
We study transport within a spatially heterogeneous one-dimensional
quantum walk with a combination of hierarchical and random barriers.
Recent renormalization group calculations for a spatially disordered
quantum walk with a regular hierarchy of barriers alone have shown
a gradual decrease in transport but no localization for increasing
(but finite) barrier sizes. In turn, it is well-known that extensive
random disorder in the spatial barriers is sufficient to localize
a quantum walk on the line. Here we show that adding only a sparse
(sub-extensive) amount of randomness to a hierarchy of barriers is
sufficient to induce localization such that transport ceases. Our
numerical results suggest the existence of a localization transition
for a combination of both, the strength of the regular barrier hierarchy
at large enough randomness as well as the increasing randomness at
sufficiently strong barriers in the hierarchy. 
\end{abstract}
\maketitle

\section{Introduction\label{sec:Intro}}

The dynamics of discrete-time quantum walks (DTQW), in particular,
their scaling in absorption and localization phenomena, is distinctly
quantum and not observed for classical walks. While localization raises
the specter of many-body phenomena observed in the tight-binding model
(which is akin to a continuous-time quantum walk), there is a form
of localization behavior in DTQW~\cite{Inui05,Falkner14a} that is
distinct in that it can arise entirely without disorder, in otherwise
perfectly homogeneous systems, for single-particle processes. Our
main focus here will be on localization that entirely stops transport
\cite{Vakulchyk17}, not just for a fraction of the wave function. 

The ability to design of quantum walks with various controllable features
(here, the strength of spatial heterogeneity and randomness) has motivated
 an expanding use of the concept. The ever increasing number of experiments
with quantum walks, discrete or continuous in time, not only indicates
the growth in technical facility to control such processes~\cite{Crespi13,Grossman04,Karski2009,Manouchehri2014,Perets08,Peruzzo1500,Preiss2015,Qiang2016,Ryan05,Sansoni12,Schreiber12,schreiber_2011a,Tang18b},
it also demonstrates the intense interest in quantum walks for their
myriad of applications in quantum information processing~\cite{Ramasesh17,alles_2011a,Asboth13,chakraborty2015randomG,Childs09,Childs13,Goyal10,kendon_2002a,kendon_2007a,kurzynski_2012a,lovett_2010a,Manouchehri08,Obuse15,OPD06,Rudinger13,Shikano10,Stefanak11,Vakulchyk17},
such as in algorithms for quantum search, optimization, and linear
algebra~\cite{Ambainis07,childs_2009b,Harrow2009,Wiebe2012}. The
corresponding classical walk problem, although far less rich in phenomenology,
has nonetheless been explored in meticulous detail over the last century
\cite{Redner01,Weiss94,Hughes96,Feller66I,Havlin87}, due to its fundamental
importance to diffusive transport as well as to randomized algorithms.
In an age dominated by synthetic nanotechnology appearing in everyday
devices, all forms of quantum transport are bound to  attain similar
importance. The basic construction of quantum walks allows for many
options that could significantly impact algorithmic performance, exemplified
by the internal degrees of freedom in coin space of DTQW, essential
to reach Grover-search efficiency in \emph{2d}~\cite{ambainis_2003a,AKR05,Boettcher18a}.
It is important to assess the robustness of the expected algorithmic
efficiency over such an array of choices, as well as to exploit these
options to control and optimize it.

The real-space renormalization group (RG) was designed as a method
to categorize the behavior of entire families of statistical process
into universality classes~\cite{Plischke94}. Based on prior applications
of RG to percolation on hierarchical networks embedded in the \emph{1d}-line
(in collaboration with Bob Ziff)~\cite{Boettcher09c,Boettcher11d},
we have extended these methods to DTQW in heterogeneous environments
with location-dependent transition operators~\cite{Boettcher20b}.
The disorder there was hierarchical  with a regular progression. While
the strength of this hierarchy systematically reduced transport, it
did not result in any localization. In contrast, in Ref.~\cite{Vakulchyk17}
it was shown with transfer matrix methods that even small amounts
of randomness at every site of a DTQW on a line can lead to localization.
Here, we show that adding a sparse amount of randomness - not on every
site but merely at every level of the hierarchy - produces an interesting
set of localization transitions by varying  a combination of both,
the barrier strength and the degree of randomness. These numerical
studies will pave the way to  precise  RG calculations in the
future.

\begin{figure*}
\vspace{-0.4cm}

\hfill{}\includegraphics[viewport=0bp 160bp 610bp 618bp,clip,width=1\textwidth]{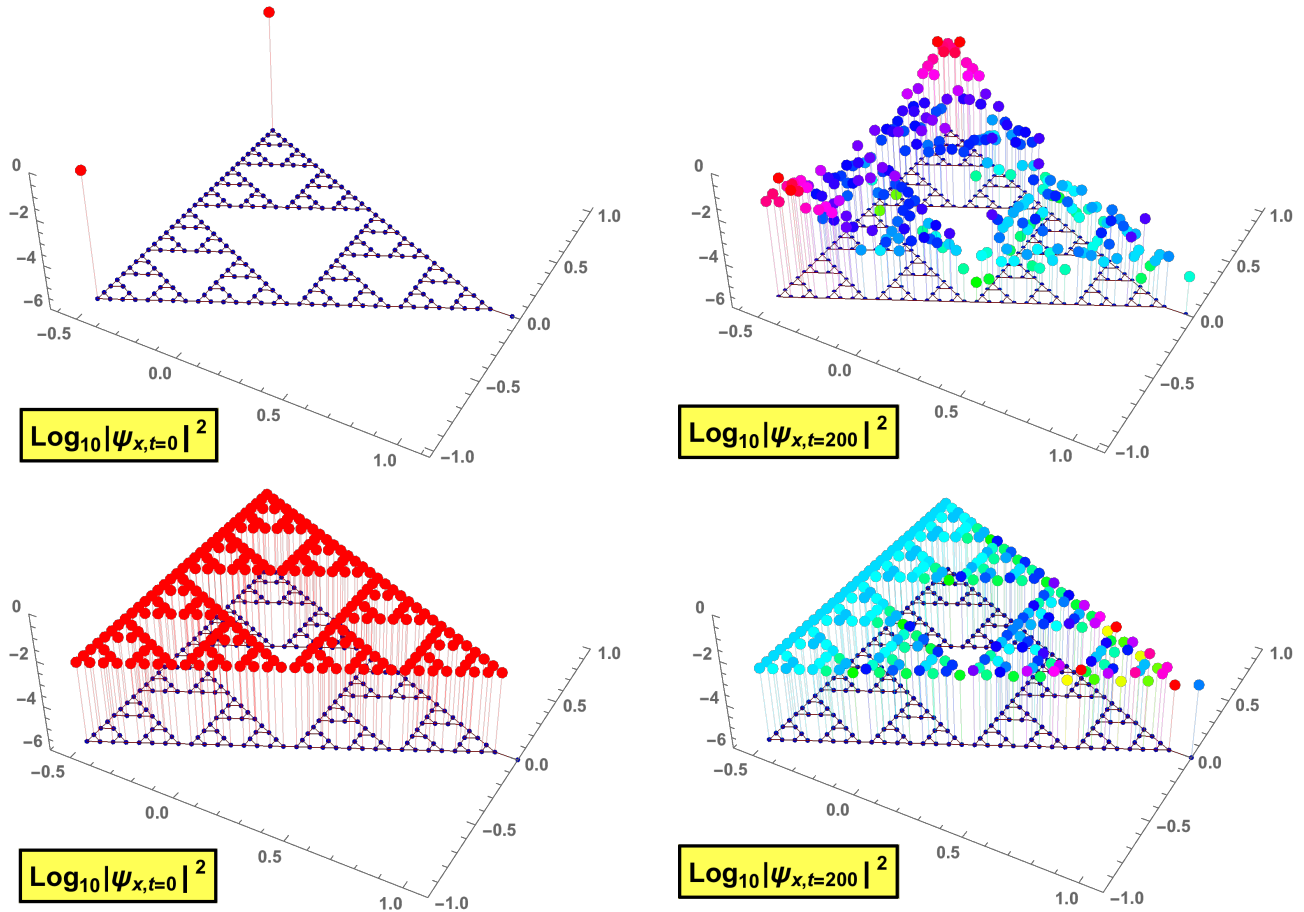}\hfill{}

\vspace{-0.8cm}

\caption{\label{fig:DSGlocalize}Numerical simulations of localization and
transport of a quantum walk in the dual Sierpinski gasket (DSG). Top
panels show a quantum walk that starts locally at $t=0$ at two corners,
with maximal separation to an exit at the $3^{{\rm rd}}$ corner (on
the right in each panel). Already at $t=200$ (for this system size,
$N=3^{5}=243$), all of the wave function $\psi_{x,t}$ (but for a
small fraction absorbed at the exit that vanishes with $N\to\infty$)
has localized throughout the system (top right panel), forever quivering,
but with essentially no further arrivals at the exit. Bottom panels
show the same process but with a uniform initial state, $\left|\psi_{x,t=0}^{2}\right|=\frac{1}{N}$
(left panel), as used for Grover's quantum search~\cite{Gro97a}.
Then, the walk remains almost uniform (right panel, for $t=200$)
and drains through the exit like water from a bathtub, without any
localization.}
\end{figure*}

A beautiful example of the connection between localization and transport
in DTQW is provided by the following observation, first mentioned
in Ref.~\cite{QWNComms13} and illustrated in Fig.~\ref{fig:DSGlocalize}.
Those simulations concern a homogeneous DTQW on a dual Sierpinski
gasket (DSG) with an absorbing wall(acting as an egress at one of
its three corners. For an initial wave function spread uniformly across
the network, no localization occurs and its entire weight gets absorbed
at the egress rapidly, similar to any classical random walk. When
DTQW is initiated while positioned at  the opposite corners, however,
the weight gets absorbed with a probability that decreases as an (as
of yet undetermined) \emph{power} of the distance between  corner
and egress. Consequently, an increasing fraction of the walk's weight
must become localized ever-farther from the starting sites. Similar
localization phenomena of quantum walks in perfectly ordered lattices
have been studied extensively~\cite{inui_2004a,IK05,Shikano10,joye_2012a,Crespi13,Ide2014}.
However, in those cases, localization is very sharp -- simply exponential
-- and relatively easily understood, as we have shown~\cite{Falkner14a}.
In contrast, the broad localization on DSG is  non-trivial and
ultimately consumes the entire walk, i.e., the arrival probability
at the wall vanishes, for diverging separations.

Our discussion is organized as follows: In Sec.~\ref{sec:DTQW}, we
start with a review of some of the fundamentals about DTQW, outline
the question about asymptotic scaling in spread (i.e., transport)
we will be concerned with, and we will recount the results for the
original \emph{1d} quantum ultra-walk without randomness from 
Ref.~\cite{Boettcher20b}. In Sec.~\ref{sec:Quantum-Walks-Randomness},
we will introduce that model extended by randomness, describe the
methods we are using to determine localization, and discuss our results.
In Sec.~\ref{sec:Conclusions}, we conclude with a summary  and provide an outlook on future work. 

\section{Discrete-Time Quantum Walks\label{sec:DTQW}}

\subsection{Evolution equation for a walk\label{subsec:Evolution-equation-for}}

Our walks are governed by the discrete-time evolution equation~\cite{Redner01}
for the state of the system, 
\begin{equation}
\left|\Psi\left(t+1\right)\right\rangle ={\cal U}\left|\Psi\left(t\right)\right\rangle \label{eq:MasterEq}
\end{equation}
with propagator ${\cal U}$. This propagator is a stochastic operator
for a classical, dissipative random walk. But in the quantum case
it is unitary and, thus, reversible. Then, in the discrete $N$-dimensional
site-basis $\left|x\right\rangle $ of some network, the PDF is given
by $\rho\left(x,t\right)=\psi_{x,t}=\left\langle x|\Psi\left(t\right)\right\rangle $
for random walks, or by $\rho\left(x,t\right)=\left|\psi_{x,t}\right|^{2}$
for quantum walks. A discrete Laplace-transform (or ``generating
function'')~\cite{Redner01} of the site amplitudes 
\begin{equation}
\overline{\psi}_{x}\left(z\right)={\textstyle \sum_{t=0}^{\infty}}\psi_{x,t}z^{t}\label{eq:LaplaceT}
\end{equation}
has all its poles -- and hence those for $\overline{\rho}\left(x,z\right)$
-- located right on the unit-circle in the complex-$z$ plane~\cite{Boettcher16}.

On the \emph{1d}-line. the propagator in Eq.\ (\ref{eq:MasterEq})
is 
\begin{equation}
{\cal U}={\textstyle \sum_{x}}\left\{ A_{x}\left|x+1\right\rangle \left\langle x\right|+B_{x}\left|x-1\right\rangle \left\langle x\right|+M_{x}\left|x\right\rangle \left\langle x\right|\right\} \label{eq:propagator}
\end{equation}
for nearest-neighbor transitions. While the norm of $\rho$ for random
walks merely requires conservation of probability for the hopping
coefficients, $A_{x}+B_{x}+M_{x}=1$, for quantum walks it demands
unitary propagation, $\mathbb{I}={\cal U}^{\dagger}{\cal U}$. The
rules~\cite{Boettcher18a} then impose the conditions $\mathbb{I}_{r}=A_{x}^{\dagger}A_{x}+B_{x}^{\dagger}B_{x}+M_{x}^{\dagger}M_{x}$
and $0=A_{x-1}^{\dagger}M_{x}+M_{x}^{\dagger}B_{x+1}=A_{x-1}^{\dagger}B_{x+1}$.
This algebra requires at least $r=2$-dimensional matrices, and it
is customary~\cite{PortugalBook,VA12} to employ the most general
unitary coin-matrix, 
\begin{equation}
{\cal C}=\left(\begin{array}{cc}
\sin\theta, & e^{i\chi}\,\cos\theta\\
e^{i\vartheta}\cos\theta, & -e^{i\left(\chi+\vartheta\right)}\sin\theta
\end{array}\right).\label{eq:QW1dCoin}
\end{equation}
We thus define ${\cal U}={\cal S}\left({\cal C}\otimes\mathbb{I}_{N}\right)$
with shift ${\cal S}$ using matrices $S^{\left\{ A,B,M\right\} }$
for transfer in a direction either out of a site or back to itself,
i.e., $A=S^{A}{\cal C}$, $B=S^{B}{\cal C}$, and $M=S^{M}{\cal C}$
with $S^{A}+S^{B}+S^{M}=\mathbb{I}_{r}$, where ${\cal C}={\cal C}_{x}$
may be heterogeneous via $x$-dependent parameters. The quantum-coin
entangles all $r$ components of $\psi_{x,t}$ and the shift-matrices
facilitate the subsequent transitions to neighboring sites. For $r=2$,
there are no self-loops ($S^{M}=0,M=0$) and we shift upper (lower)
components of each $\psi_{x,t}$ to the right (left) using projectors
$S^{A}=\left[\begin{array}{cc}
1 & 0\\
0 & 0
\end{array}\right]$ and $S^{B}=\left[\begin{array}{cc}
0 & 0\\
0 & 1
\end{array}\right]$. Alternatively, these coin-degrees of freedom could be replaced by
a ``staggered'' walk without coin~\cite{Portugal14,Portugal15},
for which schemes equivalent to the following RG can be developed.
Finally, similar considerations apply for walks on networks, such
as a fractal, except that the propagator ${\cal U}$ in Eq.~(\ref{eq:propagator})
is modified to reflect the respective Laplacian.

\subsection{Asymptotic Scaling for Walks\label{subsec:Asymptotic-Scaling-for}}

For a random walk, the probability density $\rho\left(\vec{x},t\right)$
to detect it at time $t$ at site $\vec{x}$, a distance $x=\left|\vec{x}\right|$
from its origin, obeys the  collapse with the scaling variable~$x/t^{1/d_{w}}$, 
\begin{equation}
\rho\left(\vec{x},t\right)\sim t^{-\frac{d_{f}}{d_{w}}}f\left(x/t^{\frac{1}{d_{w}}}\right),\label{eq:collapse}
\end{equation}
where $d_{w}$ is the walk-dimension and $d_{f}$ is the fractal dimension
of the network~\cite{Havlin87}. On a translationally invariant lattice
of any spatial dimension $d(=d_{f})$, it is easy to show that the
walk is always purely ``diffusive'', $d_{w}=2$, with a Gaussian
scaling function $f$, which is the content of many classic textbooks
on random walks and diffusion~\cite{Feller66I,Weiss94}. The scaling
in Eq.~(\ref{eq:collapse}) still holds when translational invariance
is broken or the network is fractal (i.e., $d_{f}$ is non-integer).
Such ``anomalous'' diffusion with $d_{w}\not=2$ may arise in many
transport processes~\cite{Havlin87,Bouchaud90,Redner01}. For quantum
walks, the only previously known value for a finite walk dimension
is that for ordinary lattices~\cite{grimmett_2004a}, where Eq.~(\ref{eq:collapse})
generically holds with $d_{w}=1$, indicating a ``ballistic'' spreading
of the quantum walk from its origin. This value has been obtained
for various versions of one- and higher-dimensional quantum walks,
for instance, with so-called weak-limit theorems~\cite{konno_2003a,grimmett_2004a,Segawa06,Konno08,VA12}.

In recent work, we have developed RG for discrete-time quantum walk
with a coin~\cite{Boettcher13a,QWNComms13,Boettcher14b,Boettcher16}.
It expands the analytic tools to understand quantum walks, since it
works for networks that lack translational symmetries. Our RG provides
principally similar results as in Eq.\ (\ref{eq:collapse}) in terms
of the asymptotic scaling variable $x/t^{1/d_{w}}$ (or pseudo-velocity~\cite{Konno05}), whose existence allows to collapse all data for
the probability density $\rho\left(\vec{x},t\right)$, aside from
oscillatory contributions (``weak limit''). While algebraically
laborious, we have developed a simple scheme to obtain RG-flow equations
for unitary evolution equations~\cite{Boettcher17a}. Abstracting
from those results, we have conjectured that the fundamental quantum
walk dimension $d_{w}$ for a homogeneous walk always is half of that
for the corresponding random walk~\cite{Boettcher14b}, 
\begin{equation}
d_{w}^{Q}=\frac{1}{2}\,d_{w}^{R}.\label{eq:dQW=00003D00003DdRW/2}
\end{equation}
It is not clear how spatial inhomogeneity affects the relation between
classical and quantum walks. The ability to explore a given geometry
much faster than diffusion is essential for the effectiveness of quantum
search algorithms~\cite{Ambainis07,childs_2003b}. In fact, using
Eq.~(\ref{eq:dQW=00003D00003DdRW/2}) and the Alexander-Orbach relation~\cite{Alexander82}, $d_{w}^{R}d_{s}=2d_{f}$, we have shown~\cite{Boettcher18a}
that attaining Grover's limit in quantum search on a homogeneous
network is determined by its spectral dimension $d_{s}$.

\subsection{Quantum Ultra-Walk \label{subsec:Strongly-inhomogeneous-quantum}}

Walks and transport efficiency in disordered environments have been
of significant interest, exemplified by the Sinai model~\cite{Sinai82}.
There have been a few approaches to understand quantum walks with
disorder, either through spatially~\cite{Brun03,Segawa06,konno_2009,Shikano10,Vakulchyk17}
or temporally~\cite{ribeiro_2004a} varying coins. Even less is known
about the impact of heterogeneous environments on quantum search efficiency.
However, exact quantum models similar to Sinai's for asymptotic scaling
of the displacement in random environments are hard to find. For random
walks, models of ``ultra-diffusion'' with a hierarchy of ultra-metric
barriers~\cite{Teitel85,Ogielski85,Huberman85,Maritan86} have been
proposed to study slow relaxation and aging, solved with
RG, that allow to interpolate between regular diffusion ($d_{w}=2$)
via anomalous sub-diffusion to the full disorder limit ($d_{w}\to\infty$).
Even spectral properties of the tight-binding model have been explored~\cite{Ceccatto87}.

\begin{figure}
\hfill{}\includegraphics[viewport=0bp 0bp 630bp 430bp,clip,width=1\columnwidth]{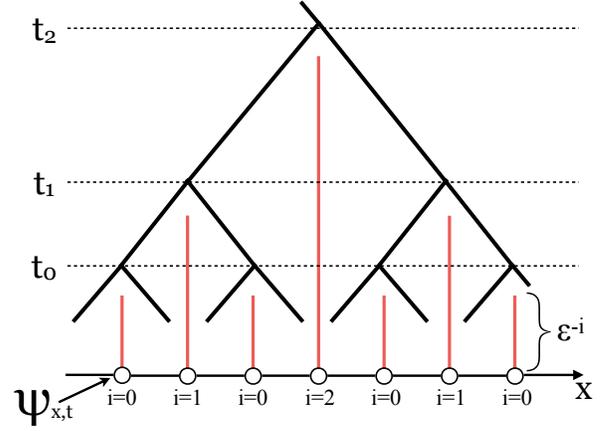}\hfill{}

\caption{\label{fig:UltraBarriers}Depiction of the hierarchical set of barriers
(red) of relative reflectivity $\epsilon^{-i}$ for $0<\epsilon<1$
and hierarchical index $i$ on a 1d-line, implemented by $\theta_{i}=\frac{\pi}{4}\epsilon^{i}$
for $0<\epsilon\protect\leq1$ in Eq.~(\ref{eq:QW1dCoin}) for the
quantum walk. These barriers generate an ultra-metrically arranged
set of domains (tree) with a hierarchy of characteristic timescales
$t_{i}$ for escape.}
\end{figure}

As a hierarchical model of spatial inhomogeneity, we have considered
position-dependent coins, Eq.~(\ref{eq:QW1dCoin}), in such a way
that all sites of odd index $x$ share the same coin, and so do all
sites that are divisible by $2^{i}$, $i\geq0$, so that sites of
the same value of $i$ have an identical coin, ${\cal C}_{i}$, with
hierarchy index $i=i(x)$ based on the (unique) binary decomposition
of any integer $x(\not=0)$~\cite{Boettcher20b}: 
\begin{equation}
x=2^{i}(2j+1),\quad(i\geq0,-\infty<j<\infty)\label{eq:hierarchyindex}
\end{equation}
Setting uniformly $\vartheta=\chi=0$ in Eq.~(\ref{eq:QW1dCoin})
but choosing 
\begin{equation}
\theta_{i}=\theta_{0}\epsilon^{i}\qquad(0<\epsilon\leq1)\label{eq:hierarchicalBarrier}
\end{equation}
with $\theta_{0}=\frac{\pi}{4}$ , the sequence of such coins becomes
ever more \emph{reflective} for a walker trying to transition through
the respective site. Thus, the walker gets confined in a tree-like
ultra-metric set of domains with vastly varying timescales for exit.
Two neighboring domains at level $i$ form a larger domain at level
$i+1$, and so on, from which an ultra-metric hierarchy emerges, as
depicted in Fig.~\ref{fig:UltraBarriers}.

The master equation (\ref{eq:MasterEq}) in Laplace-space with ${\cal U}$
in Eq.~(\ref{eq:propagator}) becomes $\overline{\psi}_{x}=zM_{x}\overline{\psi}_{x}+zA_{x}\overline{\psi}_{x-1}+zB_{x+1}\overline{\psi}_{x+1}$.
For simplicity, we merely consider initial conditions (IC) localized
at the origin, $\psi_{x,t=0}=\delta_{x,0}\psi_{IC}$ and define the
coin at $x=0$ simply to be the identity matrix. For the RG in Ref.
\cite{Boettcher20b}, we recursively eliminated $\overline{\psi}_{x}$
for all sites for which $x$ is an odd number ($i=0$), then set $x\to x/2$
($i\to i-1$) for the remaining sites, and repeat, step-by-step for
$k=0,1,2,\ldots$. In each step, we successively eliminate all sites
within an entire hierarchy $i$, each with an \emph{identical} coin
${\cal C}_{i}$, starting at $k=0$ with the ``raw'' hopping operator
$A_{i}^{(0)}=zA_{x(i,j)}$, $B_{i}^{(0)}=zB_{x(i,j)}$, and $M_{i}^{(0)}=zM_{x(i,j)}\equiv0$.
After each step, the master equation becomes \emph{self-similar} in
form by identifying the renormalized hopping operators $A_{i}^{(k)}$,
$B_{i}^{(k)}$, $M_{i}^{(k)}$ for all $i>0$ as 
\begin{eqnarray}
A_{i-1}^{(k+1)} & = & A_{0}^{(k)}\left[\mathbb{I}-M_{0}^{(k)}\right]^{-1}A_{i}^{(k)},\nonumber \\
B_{i-1}^{(k+1)} & = & B_{0}^{(k)}\left[\mathbb{I}-M_{0}^{(k)}\right]^{-1}B_{i}^{(k)},\\
M_{i-1}^{(k+1)} & = & M_{i}^{(k)}+A_{0}^{(k)}\left[\mathbb{I}-M_{0}^{(k)}\right]^{-1}B_{i}^{(k)}\label{eq:recur1dPRWmass-1}\\
 &  & \quad+B_{0}^{(k)}\left[\mathbb{I}-M_{0}^{(k)}\right]^{-1}A_{i}^{(k)}.
\end{eqnarray}
Amazingly, we can entirely eliminate the hierarchy-index $i$: If
we define the $k$-th renormalized shift matrices via $\left\{ A,B,M\right\} _{i}^{(k)}=S_{k}^{\{A,B,M\}}{\cal C}_{i+k},$
these satisfy the recursions: 
\begin{eqnarray}
S_{k+1}^{\{A,B\}} & = & S_{k}^{\{A,B\}}\left[{\cal C}_{k}^{-1}-S_{k}^{M}\right]^{-1}S_{k}^{\{A,B\}},\label{eq:Sflow}\\
S_{k+1}^{M} & = & S_{k}^{M}+S_{k}^{A}\left[{\cal C}_{k}^{-1}-S_{k}^{M}\right]^{-1}S_{k}^{B}\\
 &  & \quad+S_{k}^{B}\left[{\cal C}_{k}^{-1}-S_{k}^{M}\right]^{-1}S_{k}^{A},
\end{eqnarray}
which instead now have an explicit $k$-dependence via the inverse
coins ${\cal C}_{k}^{-1}$ of the $k$-th hierarchy.

As a specific physical situation for such a setting, Ref.~\cite{Boettcher20b}
considered a walk between two absorbing walls of separation $N=2^{l}+1$,
equidistant from the starting site $x=2^{l-1}$. As the wall-sites
$x=0$ and $x=2^{l}$ a fully absorbing, there is no flow reflecting
out of those sites such that at the end of $l-1$ RG-steps, for either
wall it is 
\begin{equation}
\overline{\psi}_{\left\{ 0,2\right\} }=S_{l-1}^{\left\{ A,B\right\} }\left({\cal C}_{l-1}^{-1}-S_{l-1}^{M}\right)^{-1}\psi_{IC}.\label{eq:AbsoPsi}
\end{equation}
In Ref.~\cite{Boettcher20b}, the RG calculation yielded for the quantum
ultra-walk the walk dimension, as defined in Eq.~(\ref{eq:collapse}):
\begin{equation}
d_{w}^{Q}=\frac{1}{2}+\frac{1}{2}\log_{2}\left(1+\epsilon^{-2}\right),\label{eq:dwQ}
\end{equation}
which grows without bound for decreasing $\epsilon$, i.e., for increasing
barrier heights transports diminishes from ballistic ($d_{w}^{Q}=1$
for $\epsilon=1$) to extremely sub-diffusive for $\epsilon\to0$.
However, it was found numerically that, eventually, the \emph{entire}
weight of the wave function gets absorbed at arbitrarily distant walls
for any finite value of $\epsilon$.

\section{Quantum Walks with Randomness\label{sec:Quantum-Walks-Randomness}}

Ref.~\cite{Vakulchyk17} has investigated the behavior of DTQW on
a 1d-line with coin parameters having extensive randomness, i.e.,
a different value for one of the variables $\theta$, $\vartheta$,
or $\chi$ in Eq.~(\ref{eq:QW1dCoin}) on every site $x$ (while the
others were held fixed throughout). Unlike for the walk with regular
hierarchical variation of $\theta$ according to Eq.~(\ref{eq:hierarchicalBarrier})
that we have described in Sec. \ref{subsec:Strongly-inhomogeneous-quantum},
it was shown there that such extensive randomness leads to a finite
localization length for the walk at any level of randomness. For example,
for $\vartheta=\chi=0$ in Eq.~(\ref{eq:QW1dCoin}) Ref.~\cite{Vakulchyk17}
selected at each site $x$ a random angle $\theta_{x}$ from a uniform
probability distribution, 
\begin{equation}
\mathcal{P}_{W}(\theta)=1/(2W),\qquad\frac{\pi}{4}-W\leq\theta\leq\frac{\pi}{4}+W,\label{eq:Wpdf}
\end{equation}
with a adjustable disorder strength $0\leq W\leq\pi$ and centered
such that each ${\cal C}_{x}$ becomes a Hadamard coin for vanishing
randomness, $W\to0$. It was found numerically that the DTQW localized
for any non-zero value of $W$ considered there.

\subsection{Quantum Ultra-Walk Model with Sub-extensive Randomness\label{subsec:QUWsubextensive}}

Clearly, overlaying this form of extensive randomness with the hierarchical
barriers analyzed in Sec. \ref{subsec:Strongly-inhomogeneous-quantum},
e.g. by replacing in Eq.~(\ref{eq:hierarchicalBarrier}) the constant
$\theta_{0}$ with a random variable $\theta_{x}$ at each site $x$
in addition to the barrier strengths $\epsilon^{-i}$, merely enhances
the already observed localization (see Fig.~\ref{fig:normal_disorder}
below). In turn, it appears that the question regarding which level
of \emph{sub-extensive} randomness might be required to induce a localization
transition is of some interest and can be conveniently studied in
the context of such a one-dimensional quantum walk. To this end, specifically,
we propose a simple two-parameter family of DTQW on the 1d-line with
a combination of regular hierarchical barriers described by $\epsilon$
in combination with \emph{hierarchical} randomness, manifested by
choosing random angles $\theta_{i}$ from the $W$-controlled distribution
in Eq.~(\ref{eq:Wpdf}) merely for each level $i=i(x)$, as defined
in Eq.~(\ref{eq:hierarchyindex}). Indeed, we find evidence for localization
transitions both for nontrivial values of $\epsilon$ for $W>0$ as
well as of $W$ for $\epsilon<1$, while there is no transition either
for $\epsilon=1$ at any value of $W$ or for $W=0$ at any $0<\epsilon\leq1$
(the case considered in Ref.~\cite{Boettcher20b}). In light of the
ability to treat this system with RG, see Sec. \ref{subsec:Strongly-inhomogeneous-quantum},
this will open the door for high-precision calculations via numerical
iteration and disorder-averaging of the (exact) RG-recursion equations
in the future.

\begin{figure}[!b]
\centering \includegraphics[viewport=6bp 0bp 396bp 255.6bp,clip,width=1\columnwidth]{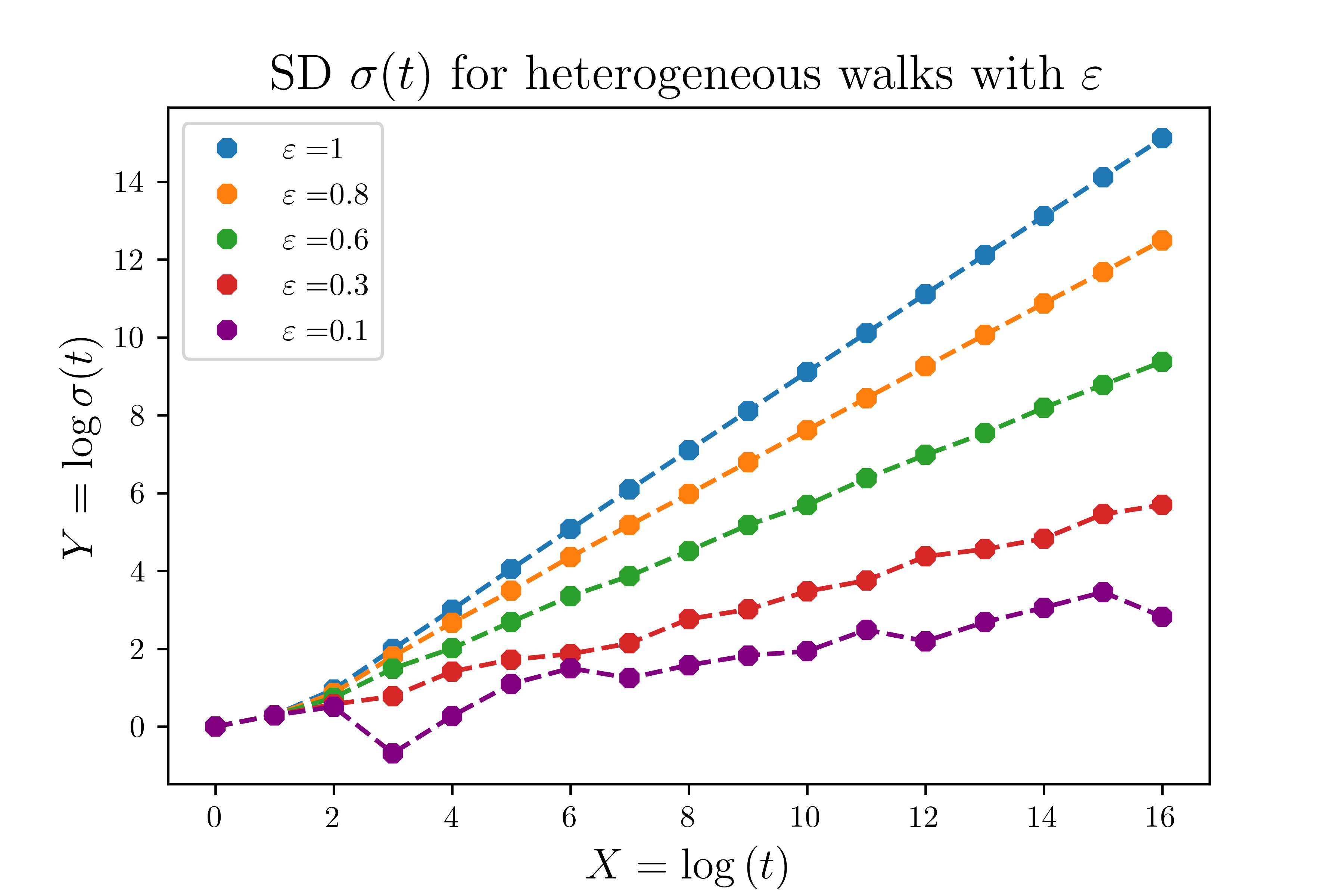}
\caption{Root mean square displacement $\sigma(t)$ for quantum ultra-walks
with varying $\epsilon$ on a logarithmic scale. It shows how the
increasing reflectivity in the hierarchy of barriers effects the transport
behavior. According to Eq.~(\ref{eq:MSD}), the slope of the curve
for each $\epsilon$ corresponds to $1/d_{w}$, whose values match
those predicted by the RG in Eq.~(\ref{eq:dwQ}) reasonably well for
all but the smallest values of $\epsilon$.}
\label{fig:spread} 
\end{figure}

\subsection{Methods\label{subsec:Methods}}

Our means to determine the existence of those transitions in this
study are very simple: For given $\epsilon$ and $W$, we generate
multiple instances of placing random angles $\theta_{i}$ drawn from
${\cal P}_{W}$ in Eq.~(\ref{eq:Wpdf}) into the coins on all sites
$x$ (up to $\left|x\right|\leq L$ with $L=2^{16}$) matching Eq.
(\ref{eq:hierarchyindex}) with that $i$ and any $\left|j\right|\leq L/2^{i+1}$.
We note that each instance involves at most a sub-extensive, $O(\log L)$
choice of random angles $\theta_{i}$ that could ever be experienced
by the walker while Ref.~\cite{Vakulchyk17} employed an extensive,
$O(L)$ selection of random angles $\theta_{x}$. For each instance,
we evolve DTQW initiated at $x=0$ for $t_{{\rm max}}=L=2^{16}$ time-steps
to measure its mean-square displacement, $\sigma(t)^{2}$, which is
the variance of $\rho(x,t)$ defined in Sec. \ref{subsec:Evolution-equation-for}.
It immediately follows from Eq.~(\ref{eq:collapse}) that the root
mean square displacement for large times $t$ scales as 
\begin{equation}
\sigma(t)\sim t^{1/d_{w}},\label{eq:MSD}
\end{equation}
from which we can extract the walk dimension $d_{w}$ asymptotically.
In Fig.~\ref{fig:spread}, we illustrate this procedure by reproducing
the walk dimensions $d_{w}$ in Eq.~(\ref{eq:dwQ}) for various values
of $\epsilon$ in the pure quantum ultra-walk without randomness ($W=0$).
For any finite value of $d_{w}$ there is still extensive transport
occurring, while $d_{w}=\infty$, or $1/d_{w}=0$, would indicate
localization. In the quantum ultra-walk, there is no localization
for any $\epsilon>0$, not even for any fraction of the wave function
\cite{Boettcher20b}.

\begin{figure}
\centering\includegraphics[viewport=20bp 85bp 350bp 520bp,clip,width=1\columnwidth]{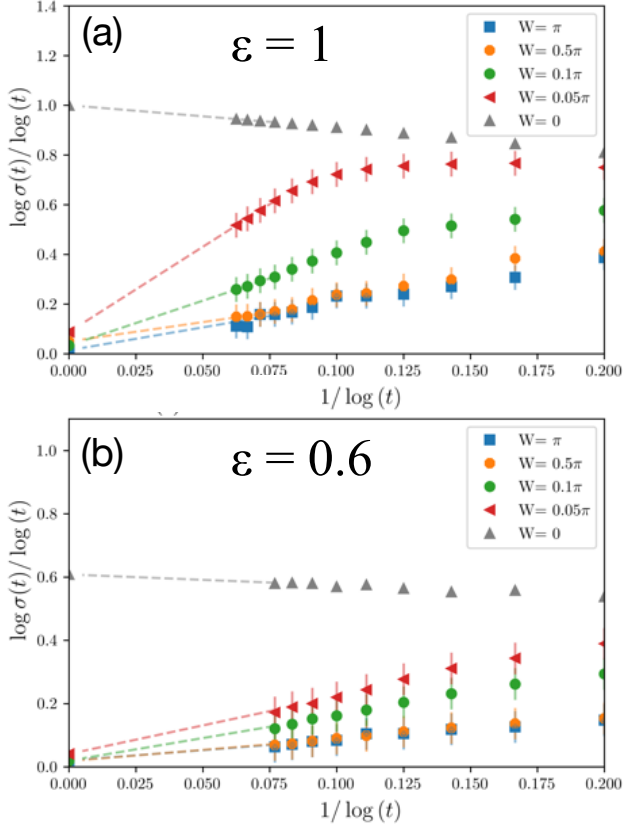}

\caption{Extrapolation plot for the root mean square displacement $\sigma(t)$
for a quantum walk with \emph{extensive} randomness at various strengths
$W$. Both for (a) $\epsilon=1$ and (b) $\epsilon=0.6$, any disorder
with $W>0$ is sufficient to drive $1/d_{w}$ to zero asymptotically,
as obtained by linear extrapolation (dashed lines) of each set of
data at the intercept with the $Y$-axis. However, as the barriers
in the system increase with decreasing $\epsilon$, the walker becomes
more readily localized. Error bars are obtained from averaging over
25 instance for $\epsilon=1$ and from averaging over 20 instances
for $\epsilon=0.6$.}
\label{fig:normal_disorder} 
\end{figure}

Since we are interested in the asymptotic behavior for large times
and distances for the walk, we instead will plot our data for $\sigma(t)$
in form of an extrapolation plot. To this end, we convert $\sigma\sim At^{\frac{1}{d_{w}}}$
in Eq.~(\ref{eq:MSD}) into 
\begin{equation}
\frac{\log{\sigma(t)}}{\log{t}}\sim\frac{1}{d_{w}}+\frac{\log A}{\log{t}}.\label{eq:extra}
\end{equation}
When plotted with $X=1/\log t$ versus $Y=\frac{\log{\sigma(t)}}{\log{t}}$,
asymptotically, Eq.~(\ref{eq:extra}) describes a line from which
we can read off $1/d_{w}$ approximately at the intercept $X=0$,
i.e., $t=\infty$. We first illustrate this technique in Fig.~\ref{fig:normal_disorder}
for simulations employing extensive randomness which recap the findings
of Ref.~\cite{Vakulchyk17} (for $\epsilon=1$) and show that localization
only gets stronger for higher barriers ($\epsilon=0.6)$. Clearly,
in all cases with $W>0$, the extrapolations indicate a vanishing
of $1/d_{w}$, making $\sigma(t)$ bounded for large times $t$ as
evidence that the walk remains localized.

\subsection{Results\label{subsec:Results}}

\begin{figure}
\centering \includegraphics[viewport=10bp 20bp 350bp 620bp,clip,width=1\columnwidth]{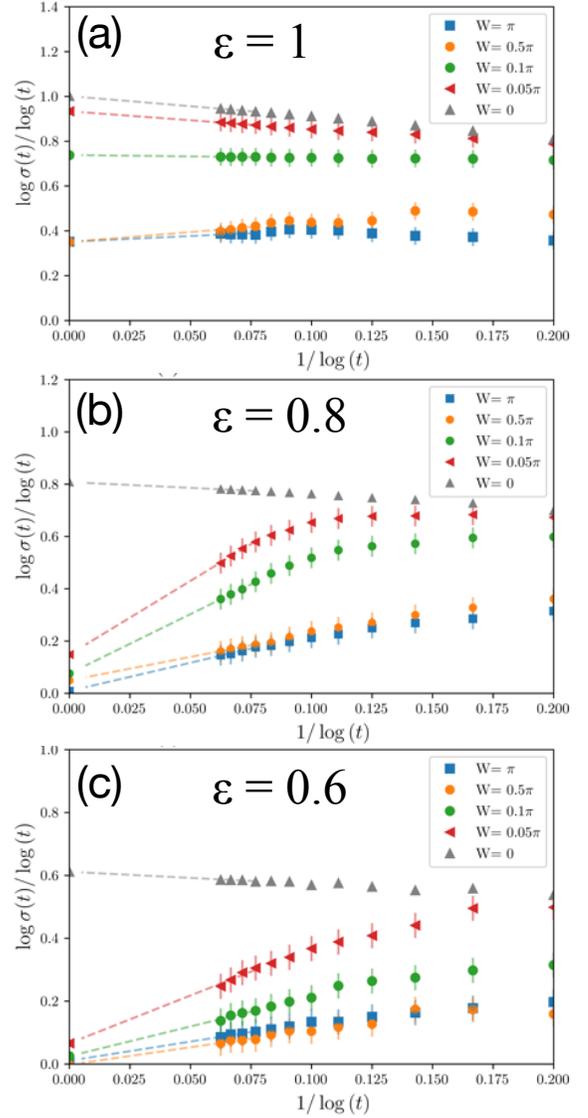}

\caption{Extrapolation plots for the quantum ultra-walk model with various
strengths $W$ of sub-extensive randomness at a fixed barrier strength
of (a) $\epsilon=1$, (b) $\epsilon=0.8$, and (c) $\epsilon=0.6$.
For $\epsilon=1$, the case of a homogeneous quantum walk, the addition
of sub-extensive randomness of any strength $W$ is insufficient to
bring $1/d_{w}$ even anywhere near to vanishing. For lower $\epsilon$,
i.e., exponentially increasing barriers, some level of such randomness
$W>0$ does prove sufficient to induce an apparent localization transition
where there was none without $(W=0)$. Note that for lower $\epsilon$
(i.e., stronger barriers), the same amount of randomness suppresses
transport more. Error bars were obtained by averaging over 50 instances.}
\label{fig:novel_disorders}
\end{figure}

In the following, we apply the methods developed in Sec. \ref{subsec:Methods}
to the model of a quantum ultra-walk with sub-extensive randomness
introduced in Sec. \ref{subsec:QUWsubextensive}. In Fig.~\ref{fig:novel_disorders},
we summarize our data of the walk simulations for various values in
the ($\epsilon,W$)-plane. In particular, as shown in Fig.~\ref{fig:novel_disorders}(a),
for the otherwise homogeneous (barrier-free) \emph{1d }quantum walk
obtained for $\epsilon=1$, the addition of merely sub-extensive randomness
placed hierarchically on the lattice is insufficient to localize the
walk for any value of $W$. Even for angles chosen entirely randomly
($W=\pi)$ the walk at most becomes mildly sub-diffusive and $1/d_{w}$
remains far from zero. This is in stark contrast with the corresponding
case of extensive randomness~\cite{Vakulchyk17} shown in Fig.~\ref{fig:normal_disorder}(a).
However, in concert with the hierarchy of barrier that emerges for
$\epsilon=0.8$ and $\epsilon=0.6$, as shown in Figs. \ref{fig:novel_disorders}(b,c),
such a sub-extensive amount of randomness proves sufficient to induce
localization. In fact, it appears that for each of those fixed values
of $\epsilon<1$, there is a transition at a finite value of $W$,
although it would be possible for that transition to be at $W=0$. 

\begin{figure}
\centering\includegraphics[viewport=0bp 40bp 725bp 550bp,clip,width=1\columnwidth]{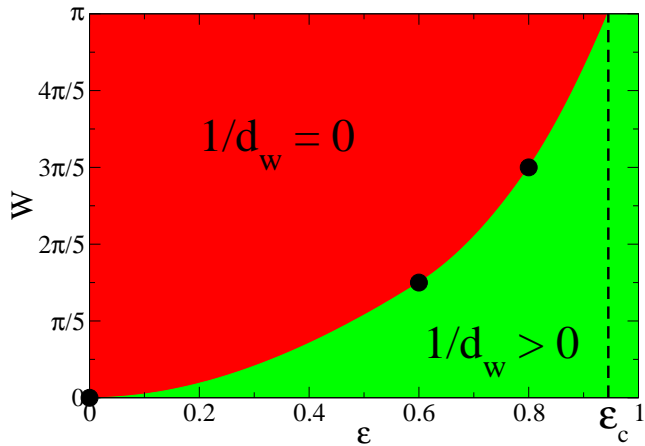}\caption{\label{fig:epsWphase}Sketch of the proposed phase diagram for the
localization transition in the $(\epsilon,W)$-plane for the quantum
ultra-walk model on a line with randomness. Eq.~(\ref{eq:MSD}) suggests
that walks with $1/d_{w}=0$ (red) become localized, while those with
$1/d_{w}>0$ (green) maintain transport. The phase line is estimated
based on the fact (1) that Eq.~(\ref{eq:dwQ}) provides $1/d_{w}>0$
for all $\epsilon>0$ without randomness ($W=0$) while $1/d_{w}=0$
at $\epsilon=0$, (2) that Figs. \ref{fig:novel_disorders}(b-c) show
such a transition at intermediate values of $W$ for $\epsilon=0.6$
and $0.8$ (black dots), and (3) that Fig.~\ref{fig:novel_disorders}(a)
shows no transition for any $W$ when $\epsilon=1$. The strength
of transport for $\epsilon=1$ even at maximum (sub-extensive) randomness,
$W=\pi$, as apparent in Fig.~\ref{fig:novel_disorders}(a), suggests
that the localized domain terminates at some $\epsilon_{c}<1$. }

\end{figure}

We can summarize our results by sketching out a tentative phase diagram
for the localization transition in the plane formed by the set of
parameters $(\epsilon,W)$. In Fig.~\ref{fig:epsWphase}, we illustrate
the emerging scenario, highlighting the region of localization ($1/d_{w}=0$,
in red) from that of transport ($1/d_{w}>0$, in green) with a phase
boundary between them estimated from our data. First, the earlier
discussion in Sec. \ref{subsec:Strongly-inhomogeneous-quantum} concerning
the quantum ultra-walk without any randomness shows that the entire
line ($\epsilon,W=0$) is not localized but that the line ($\epsilon=0,W$)
for any $W$ certainly is. Thus, the phase boundary must pass the
origin ($\epsilon=0,W=0$). While we can not entirely exclude the
possibility that that boundary remains at $W=0$ also for most $\epsilon<1$,
it is conceivable from Figs. \ref{fig:novel_disorders}(b,c) that
for intermediate barrier strengths $\epsilon$ there is a transition
at similarly intermediate values $W$, as the two interior marks at
$\epsilon=0.6$ and $\epsilon=0.8$ insinuate. The argument for this
scenario is strengthened by the fact that the point ($\epsilon=1,W=\pi$),
yielding $1/d_{w}\approx0.3-0.4$ from Fig.~\ref{fig:novel_disorders}(a),
appears far from localization and, thus, from the phase boundary.
Hence, unless there is a discontinuous jump in $d_{w}$, that phase
boundary seems to reach full randomness ($W=\pi$) at some value of
$\epsilon_{c}$ that might be close to, but is well bounded-away from
$\epsilon=1$, as indicated in Fig.~\ref{fig:epsWphase}. Then, either
the phase boundary varies gradually between the origin and that point,
as shown, or discontinuously jumps at $\epsilon_{c}$ from $W=0$
to $W=\pi$.

\section{Conclusions\label{sec:Conclusions}}

Based on the recent solution of the quantum ultra-walk~\cite{Boettcher20b},
a \emph{1d} DTQW that evolves through a spatially heterogeneous (albeit
not random) environment characterized by a hierarchy of progressively
diverging barriers (in the form of increasingly reflective quantum
coins), we have extended this model by introducing a certain, sub-extensive
amount of spatial randomness. The effect of extensive randomness,
with random coin parameters on every site $x$, on a \emph{1d} DTQW
has been well-studied and shown to lead to localization for even the
smallest amount of randomness~\cite{Vakulchyk17}. Our new model permitted
us to explore the impact of sub-extensive randomness on localization
as well as to trace out an interesting phase diagram with regions
of localization and transport separated by a phase boundary, generated
by the interplay of such randomness with the hierarchy of barriers.
We have shown that, while each ineffective by themselves, such sparse
randomness in combination with even mildly escalating barriers readily
induces localization. While not yet resolved in great detail, the
available evidence already suggests a few interesting features exhibited
within that parameter space, with localization transitions occurring
at either a non-trivial finite randomness or finite barrier strength,
or both. Aside from the myriad applications of highly controllable
DTQW in quantum algorithms~\cite{Childs09,Childs13}, this model provides
also a simple example with the potential for such a complex phase
diagram. It would be interesting to supplement these studies with
a similar continuous-time quantum walk~\cite{LiBo16}. While likely
equivalent in most aspects~\cite{Strauch06}, we have favored a discrete-time
formulation here, since the extra coin-space allows for richer designs
in the walk dynamics.

In future work, we intend to employ RG to gain a more precise description
of this diagram. While it is straightforward to obtain the exact RG-recursions
from Eq.~(\ref{eq:Sflow}) that could be evolved numerically, replacing
the regular progression of hierarchical barriers solved in Ref.~\cite{Boettcher20b}
with a random sequence makes the prospect of receiving analytical
results (even asymptotically) very daunting. Since contributions to
the fixed points of the RG seem to arise from many poles~\cite{Boettcher16}
throughout the complex-$z$ plane in Laplace space, originating in
Eq.~(\ref{eq:LaplaceT}), even a numerical evolution of the RG-recursions
does not yield insights easily. Alternatively, these questions should
be more readily accessible using numerically the transfer matrix method
of Ref.~\cite{Vakulchyk17}.

\subsubsection*{Acknowledgements}

This work is but a small contribution to acknowledge the tremendous
impact that Bob Ziff, by strength of his kindness and intellect, has
on all who know him and on the field of statistical physics. 

\bibliographystyle{apsrev4-1}
\bibliography{/Users/sboettc/Boettcher}

\end{document}